\documentclass[aps,reprint,nofootinbib,superscriptaddress,longbibliography]{revtex4-1}

\usepackage[english]{babel}
\usepackage[utf8x]{inputenc}
\usepackage[T1]{fontenc}

\usepackage{float}
\usepackage{amsmath}
\usepackage{amssymb}
\usepackage[colorlinks]{hyperref}
\usepackage{cleveref}
\usepackage{booktabs}
\usepackage{graphicx}
\usepackage{glossaries}

\usepackage{physics}
\usepackage{siunitx}
\usepackage{array}
\usepackage{tabularx}

\newcommand{\delay}[1]{\mathbf{D}_{#1}}

\newcommand{\conj}[1]{\opbraces{{#1}^\ast}}
\newcommand{\psd}[1]{\opbraces{S_{#1}}}

\renewcommand{\qc}{\,\text{,}}
\newcommand{\qs}{\,\text{.}}

\DeclareSIUnit\solarmass{M_\odot}

\newacronym{lisa}{LISA}{Laser Interferometer Space Antenna}
\newacronym{gw}{GW}{gravitational-wave}
\newacronym{gpu}{GPU}{graphics processing unit}
\newacronym{esa}{ESA}{European Space Agency}
\newacronym{tdi}{TDI}{time-delay interferometry}
\newacronym{fir}{FIR}{finite impulse response}
\newacronym{iir}{IIR}{infinite impulse response}
\newacronym{psd}{PSD}{power spectral density}
\newacronym{asd}{ASD}{amplitude spectral density}
\newacronym{ldc}{LDC}{LISA Data Challenge}
\newacronym{ssb}{SSB}{Solar system's barycenter}
\newacronym[longplural=moveable optical sub-assemblies]{mosa}{MOSA}{moveable optical sub-assembly}
\newacronym{mcmc}{MCMC}{Markov chain Monte Carlo}
\newacronym{snr}{SNR}{signal-to-noise ratio}
\newacronym{cpu}{CPU}{central processing unit}
\newacronym{emri}{EMRI}{extreme mass-ratio inspiral}
\newacronym[longplural=maxima a posteriori]{map}{MAP}{maximum a posteriori}

\begin{document}

\title{Assessing the data-analysis impact of LISA orbit approximations using a GPU-accelerated response model}

\author{Michael L. Katz}
\email{michael.katz@aei.mpg.de}
\affiliation{Max-Planck-Institut f\"ur Gravitationsphysik, Albert-Einstein-Institut, 
Am M\"uhlenberg 1, 14476 Potsdam-Golm, Germany}

\author{Jean-Baptiste Bayle}
\affiliation{Jet Propulsion Laboratory, California Institute of Technology, Pasadena CA 91109, USA}

\author{Alvin J. K. Chua}
\affiliation{Department of Physics, National University of Singapore, Singapore 117551}
\affiliation{Department of Mathematics, National University of Singapore, Singapore 119076}
\affiliation{Theoretical Astrophysics Group, California Institute of Technology, Pasadena, CA 91125, USA}

\author{Michele Vallisneri}
\affiliation{Jet Propulsion Laboratory, California Institute of Technology, Pasadena CA 91109, USA}

\date{\today}


\begin{abstract}
The analysis of \gls{gw} datasets is based on the comparison of measured time series with theoretical templates of the detector's response to a variety of source parameters. For the \gls{lisa}, the main scientific observables will be the so-called \gls{tdi} combinations, which suppress the otherwise overwhelming laser noise. Computing the \gls{tdi} response to \glspl{gw} involves projecting the \gls{gw} polarizations onto the \gls{lisa} constellation arms, and then combining projections delayed by a multiple of the light propagation time along the arms. Both computations are difficult to perform efficiently for generic \gls{lisa} orbits and \gls{gw} signals. Various approximations are currently used in practice, e.g., assuming constant and equal armlengths, which yields analytical \gls{tdi} expressions that are essential to the desirable speed of current analysis codes. In this article, we present \texttt{fastlisaresponse}, a new efficient GPU-accelerated code designed to perform systematics studies on the \gls{lisa} response in a fully Bayesian context. We examine loud Galactic binary signals first with the typical equal-armlength approximation and then with a hybrid template that uses accurate orbits for the projections and equal-armlength orbits for the \gls{tdi} combinations. The hybrid template is an attempt to preserve the efficient analytical \gls{tdi} expressions. We conclude that all equal-armlength parameter-estimation codes, including when only used for TDI, need to be upgraded to the generic response if they are to achieve optimal accuracy for high (but reasonable) \acrshort{snr} sources within the actual \gls{lisa} data.
\end{abstract}

\glsresetall
\maketitle

\section{Introduction}

The launch of the \gls{lisa} in the early 2030s will extend the reach of \gls{gw} astronomy to the millihertz band, complementing the plentiful observations of compact binaries obtained by ground-based \gls{gw} detectors~\cite{LIGOScientific:2018mvr, LVK2018LivingReview}. \gls{gw} sources in the millihertz band include Galactic binaries, typically consisting of two white dwarfs emitting quasi-monochromatic \gls{gw}s; massive--black-hole binaries, producing loud signals that sweep across the band as the black holes inspiral and merge; stellar-origin black-hole binaries, which are detected by LISA at large orbital separations and will eventually merge in the band of ground-based detectors; and extreme mass-ratio inspirals, consisting of a stellar-mass compact object orbiting a massive black hole. The combination of observations from all these sources will provide immense scientific return~\cite{LISAMissionProposal}. 

Analyzing the \gls{lisa} data stream is a tall task. Detector noise is expected to be non-stationary due to the presence of glitches, data gaps, and drifting noise levels in the \gls{lisa} components 
\cite{Edlund:2005cg, Baghi:2019eqo, Cornish:2020odn, Dey:2021dem, Katz:2022izt, 2020SciPy-NMeth,Digman:2022jmp}. Furthermore, a large number of overlapping \gls{gw} signals will be present for the duration of the observation. Therefore, estimating detector noise will be more complicated than in the ground-based case~\cite[e.g.][]{Hartwig:2021dlc, Wang:2022sti, Digman:2022jmp}, where signal-free stretches of data can be used to characterize noise directly. An additional complication is the motion of the \gls{lisa} spacecraft: the compact-binary coalescence signals observed from the ground are short enough for detectors to be assumed static, allowing for a detector response that is a function of frequency only. By contrast, the movement of the \gls{lisa} constellation in its Solar orbit leads to a response function that depends on both frequency and time~\cite{Cornish:2002rt,Cornish:2003tz, Vallisneri:2004bn,Marsat:2018oam}.

It is useful to decompose the \gls{lisa} \gls{gw} response function in two stages: in the first, we project the \gls{gw} polarizations onto the evolving \gls{lisa} arms, computing projections along all six interferometric links (forward and backward along each arm); in the second, the projections are combined with appropriate time shifts to form \gls{tdi} observables, which suppress the laser noise that would otherwise drown the \gls{gw} signals~\cite{Tinto:1999yr, Tinto:2002de, Tinto:2014lxa}. For this reason, the \gls{lisa} sensitivity is typically quoted for \gls{tdi} observables~\cite{LISAMissionProposal}.
Both stages require knowledge of the \gls{lisa} orbits: the projections involve the spacecraft positions and the light-propagation times, or delays, along all six links, while \gls{tdi} requires sufficiently accurate delays for proper laser-noise suppression.

Many \gls{lisa} analysis codes make the simplifying assumption of \emph{equal-armlength} orbits~\cite[e.g.][]{Cornish:2007if, Armstrong1999,Marsat:2020rtl,Katz:2020hku,Dey:2021dem, Katz:2021uax, Strub:2022upl, Cornish:2021smq, Boileau:2021kiw, Shuman:2021ruh, Karnesis:2021tsh, Robson:2018svj}. This unphysical model neglects the breathing of the armlengths in actual orbits~\cite[e.g.][]{Wang:2020fwa}, as well as the orbital corrections required to realign the spacecraft to counter accumulating drifts. The equal-armlength model has a period of one year and is completely determined by two parameters.
The computation of \gls{tdi} expressions is especially convenient because fixed equal armlength delays allow the resummation of single-arm responses using trigonometric identities over \gls{gw} phasing terms~\cite[see, e.g.,][]{Marsat:2018oam}.
A more realistic but still unphysical model places the spacecrafts on eccentric Keplerian orbits~\cite{Baghi:2022ucj}. It also has a period of one year and is completely determined by three parameters.

The actual \gls{lisa} orbits are influenced by the gravitational attraction of the planets, as well as non-gravitational effects such as solar wind~\cite{Frank:2020tjx} or orbital correct thrusts; thus they display drifts leading to unequal and evolving armlengths\footnote{Note that analytic Keplerian orbits also yield unequal and evolving armlengths. Variation of armlengths in the case of realistic orbits is an order of magnitude larger.}, and non-periodic trajectories. To account for these effects fully, \gls{tdi} observables must be constructed by evaluating single-arm \gls{gw} projections at delayed times that are changing continually along the orbits~\cite[e.g.][]{Vallisneri2005,Vallisneri:2020otf}.
In this work, we present the first efficient implementation on a \gls{gpu} of this \emph{general} \gls{gw} response for arbitrary \gls{lisa} orbits. 
This new code can be used directly in \gls{gw} searches and parameter estimation, avoiding any orbit or delay approximation that may affect the accuracy of results.

Conversely, the code enables an exploration of the effects of orbit approximations in \gls{lisa} data analysis. Initial studies on this topic have been performed during the early stages of LISA development~\cite{Rubbo:2003ap, Cornish:2003tz}. Using Galactic-binary waveforms, currently available settings for the LISA mission, new orbits provided directly by the \gls{esa} and our \gls{gpu} code as \emph{truth}, we perform a systematics study to further understand the actual loss of parameter-estimation accuracy incurred by computing TDI observables with the equal-armlength approximation: for moderately loud signals, we find significant bias in the recovery of important astrophysical parameters such as frequency, frequency derivative, and sky position.
We also test a hybrid approximation that uses realistic orbits for projections, and equal-armlength orbits for \gls{tdi}. The purpose of this hybrid template is to understand if we can maintain the efficiency of current analysis codes that heavily rely on the analytic \gls{tdi} functions that come from assuming equal armlengths. While such hybrid templates fit the truth much better, we still find consequential bias for louder (but realistic) sources.
Therefore, it is our recommendation that \gls{lisa} analysis codes used for actual data should fully model the effects of the \gls{lisa} orbits. Tools such as our \gls{gpu} code will ensure that the additional detail does not pose an undue burden on computation.

This paper is organized as follows: in \cref{sec:bayes}, we describe the statistical methods used in \gls{gw} analysis with \gls{lisa}; in \cref{sec:template}, we introduce the waveform templates used in our study; in \cref{sec:response}, we discuss the formulation and implementation of the response function; in \cref{sec:posterior}, we describe the results of our analysis; in \cref{sec:discuss}, we discuss the implications of our findings; last, in \cref{sec:conclude} we present our conclusions.

\section{Bayesian analysis on gravitational-wave signals}
\label{sec:bayes}

\subsection{Likelihood function}

The analysis of \gls{gw} signals is generally performed within a Bayesian framework, driven by Bayes' rule,
\begin{equation}
	p(\vec{\Theta}|d, \Lambda) = \frac{p(d | \vec{\Theta}, \Lambda)p(\vec{\Theta}|\Lambda)}{p(d |\Lambda)}
	\qc
\label{eq:bayes}
\end{equation}
where $d=d(t)$ is the measured \gls{gw} data; $\vec{\Theta}$ is the vector of parameters representing a \gls{gw} source, and is associated with the assumed model of the signal, $\Lambda$; $p(\vec{\Theta}|d, \Lambda)$ is the posterior probability distribution on the parameters of the source; $p(d | \vec{\Theta}, \Lambda)=\mathcal{L}$ is the probability that the data is represented by the chosen model and model parameters, also called the likelihood; $p(\vec{\Theta}|\Lambda)$ is the prior probability on the parameters; and $p(d |\Lambda)$ is the integral of the numerator over all parameter space, also referred to as the evidence. In this study, we will produce the posterior distribution by drawing samples from it with \gls{mcmc} techniques. Typically, the evidence is intractable in the \gls{gw} setting. With \gls{mcmc} methods, the evidence enters as a constant factor and can, therefore, be neglected. 
In order to determine the posterior, we must compute the likelihood. Under the assumption of stationary and Gaussian noise, we write down the noise-weighted inner product between two time domain signals $a(t)$ and $b(t)$,
\begin{equation}
	 \braket{a}{b} = 4 \Re \int_0^\infty{ \frac{\conj{\tilde{a}(f)} \tilde{b}(f)}{\psd{n}(f)} \dd{f}}
	 \qc
\label{eq:inner}
\end{equation}
where $\tilde{a}(f)$ is the Fourier Transform of $a(t)$ and $S_n(f)$ is the one-sided \gls{psd} of the noise. We use the ``SciRDv1'' curve from the \texttt{tdi} package from the \gls{ldc}~\cite{SciRD1} for the \gls{psd}. The log-likelihood is a linear combination of inner products,
\begin{align}
\begin{split}
	\log{\mathcal{L}} \propto -\frac{1}{2}& \braket{d - h}{d-h}
	\\
	= &-\frac{1}{2} \qty(\braket{d}{d} + \braket{h}{h} - 2 \braket{d}{h})
	\qc
\end{split}
\label{eq:like}
\end{align}
where $h(t)$ is the template that models the true signal, $s(t)$. The true signal together with the noise contribution, $n(t)$, makes the data stream: $d(t)=s(t) + n(t)$. For all studies in this work, we use no additive noise ($n(t)=0$). This allows for the direct analysis of the likelihood surface without any shift due to the random noise process. 

The optimal \gls{snr} $\rho_\text{opt}$ achievable for a template is $\sqrt{\braket{h}{h}}$. When fitting a template against the data (and assuming the true signal is perfectly modeled by the template), the \gls{snr} will be normally distributed about the optimal \gls{snr} with variance equal to 1. The extraction \gls{snr} $\rho_\text{ex}$ is given by
\begin{equation}
    \rho_\text{ex} = \frac{\braket{d}{h}}{\sqrt{\braket{h}{h}}}
    \qs
\label{eq:extract}
\end{equation}
The extraction \gls{snr} is fundamental to signal processing in terms of  understanding the detection of a signal in a time series. We use it in this work as a diagnostic indicator to help understand detection capabilities of different template models. Assuming no additive noise, the overlap (normalized cross-correlation) between the template and data is given by $\rho_\text{ex}/\sqrt{\braket{s}{s}}$. The mismatch is then defined as 1 minus the overlap.

An additional statistic that is of use in this work is the \textit{fitting factor}~\cite[e.g.][]{Owen:1998dk}. The fitting factor is the overlap between the data and the template whose parameters are located at the global maximum in the likelihood surface, which is found using stochastic sampling methods (discussed below).

\subsection{Markov chain Monte Carlo sampler}
\label{sec:mcmc}

We use the \gls{mcmc} technique to draw samples from the posterior distribution. The sampler is a combination\footnote{The sampler code is available upon request to the authors. It will soon be made publicly available.} of the overall architecture of \texttt{emcee}~\cite{emcee} with the parallel tempering methods from \texttt{ptemcee}~\cite{Vousden2016}. In all of the sampling runs used in this work, four temperatures are used ranging from 1 (target distribution) to $\infty$ (prior). Each temperature contains 50 walkers. The only proposal used was the ``Stretch'' proposal~\cite{Goodman2010}. 
 
\section{Template generation}
\label{sec:template}

As the Markov chain evolves, templates are generated at each step to evaluate the likelihood for all walkers. The first step in this process is to generate $h_+^\text{SSB}(t)$ and $h_\times^\text{SSB}(t)$, the two polarization time series at the \gls{ssb}. For all tests in this paper, we focus on single Galactic white dwarf binary, whose time domain waveform in its source frame (scaled by the distance and denoted with an ``S'' superscript) is given by
\begin{subequations}
\begin{align}
    h_+^\text{S} &= A \cos(2 \pi \qty(f_0t + \frac{1}{2}\dot{f}_0t^2  + \frac{1}{6}\ddot{f}_0t^3) + \phi_0)
    \qc
    \\
    h_\times^\text{S} &= A \sin(2 \pi \qty(f_0t + \frac{1}{2}\dot{f}_0t^2  + \frac{1}{6}\ddot{f}_0t^3) + \phi_0)
    \qc
\end{align}
\end{subequations}
where $A$ is the amplitude of the \gls{gw} at the \gls{ssb}; $\{f_0, \dot{f}_0, \ddot{f}_0\}$ are the initial \gls{gw} frequency and its first two time derivatives; and $\phi_0$ is the initial phase of the \gls{gw}. The amplitude $A$ is given by
\begin{equation}
	A(\mathcal{M}_c, f_0, d_L) = 2 \frac{(G\mathcal{M}_c)^{5/3}}{c^4 d_L}(\pi f)^{2/3} 
	\qc
\label{eq:amplitude}
\end{equation}
where $\mathcal{M}_c$ is the chirp mass and $d_L$ is the luminosity distance.
Assuming that \gls{gw} emission solely drives the evolution of the binary, the frequency derivative $\dot{f}_\mathrm{GW}$ is given by
\begin{equation}
	\dot{f}(\mathcal{M}_c, f) = \frac{96}{5}\pi^{8/3}\left(\frac{G\mathcal{M}_c}{c^3}\right)^{5/3}f^{11/3} \qs
\label{eq:fdot}
\end{equation}
For a given value of $f_0$, choosing the initial frequency derivative $\dot{f}_0$ effectively fixes the chirp mass of the binary. With $f_0$ and $\dot{f}_0$ set, we can deduce $\ddot{f}_0$ assuming once again that gravitational radiation is the sole driver of the binary evolution,
\begin{equation}
    \ddot{f}_0 = \frac{11}{3}\frac{\dot{f}_0^2}{f_0} \qs
\end{equation}
Further derivatives have very small effects and are ignored here.

These source-frame \gls{gw} orientations are then transformed to the \gls{ssb} frame using the orbital inclination $\iota$ and polarization angle $\psi$,
\begin{align}
    \mqty[h_+^\text{SSB} \\ h_\times^\text{SSB}] &= 
    \mqty[
        \cos{2\psi} & -\sin{2\psi} \\
        \sin{2\psi} & \cos{2\psi}]
    \mqty[
        -(1 + \cos^2{\iota}) h_+^\text{S} \\
        -(2 \cos{\iota}) h_\times^\text{S}
    ] \qs
\end{align}

The parameter set used above, $\{A, f_0, \dot{f}_0, \phi_0, \iota, \psi\}$, combined with the two sky-localization angles, give the eight parameters needed to generate our Galactic binary waveform. The sky-localization angles represent the ecliptic longitude $\lambda$ and latitude $\beta$. These angles will appear in the detector response described below.

\section{Instrument Response}
\label{sec:response}

\begin{figure}
    \centering
    \includegraphics[width=\columnwidth]{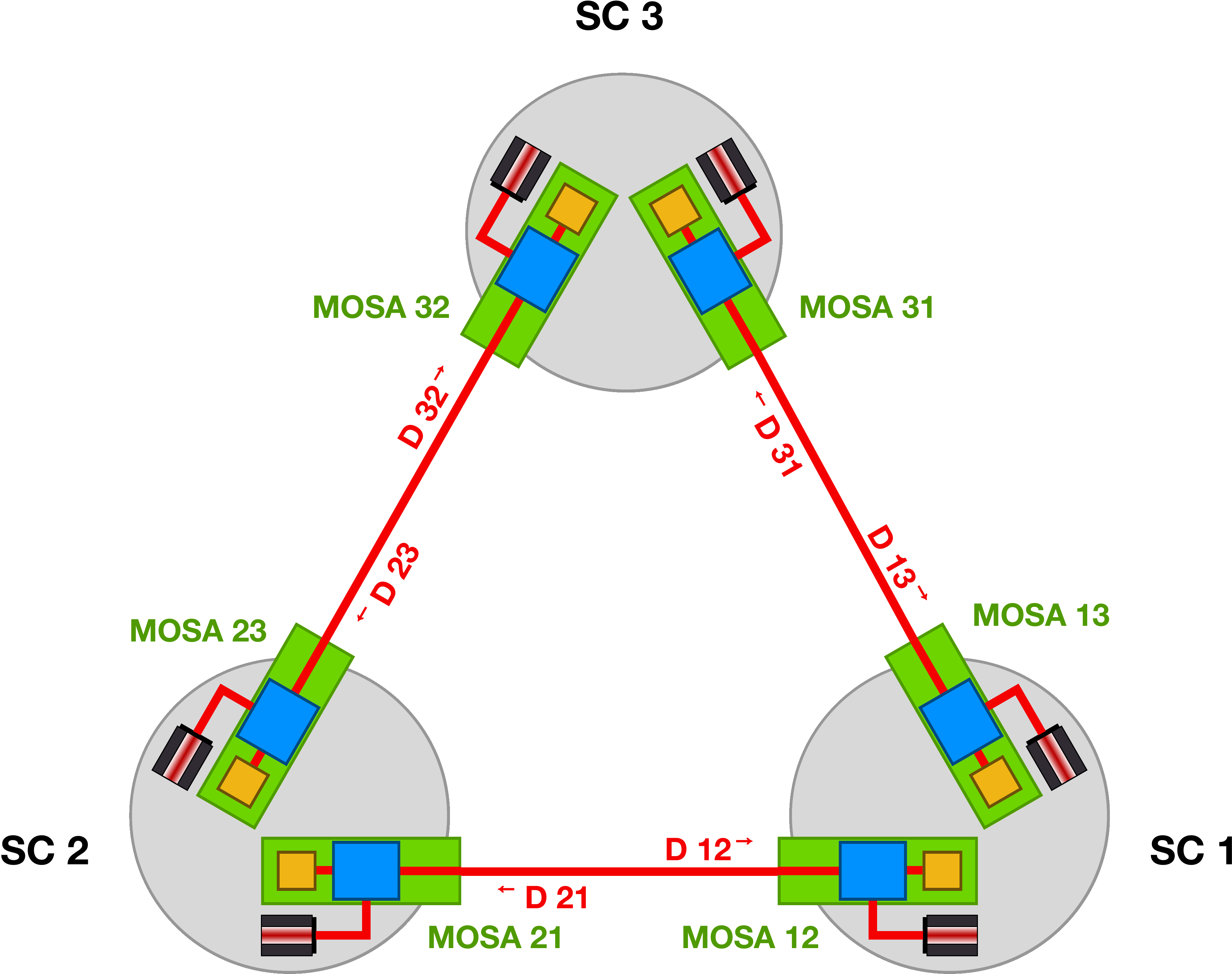}
    \caption{Indexing conventions.}
\label{fig:indexing}
\end{figure}

We follow the standard conventions given in \cref{fig:indexing}. Spacecraft are indexed from 1 to 3 clockwise when looking down on the $z$-axis. \Glspl{mosa} are indexed with two numbers $ij$, where $i$ is the index of the spacecraft the system is mounted on (local spacecraft), and $j$ is the index of the spacecraft the light is received from (distant spacecraft).

Measurements are indexed according to the \gls{mosa} they are performed on. Light travel times are indexed according to the \gls{mosa} they are measured on, i.e., the receiving spacecraft. All equations in this document possess the symmetries of the triangular constellation, i.e., 3 reflections and 3 rotations. All associated transformations on the indices can be generated from one circular permutation of indices and one reflection. Therefore, we will only give one expression and leave the reader to form all remaining expressions using this set of index transformations.

\subsection{Reference frames}

We present here the formulation of the \gls{lisa} response to gravitational waves, following the conventions proposed by the LDC Manual (available at \url{https://lisa-ldc.lal.in2p3.fr}).

\begin{figure}
    \centering
  
        
        
        
         
        
    \includegraphics[width=0.75\columnwidth]{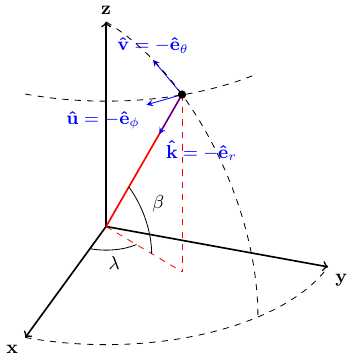}
    \caption{Source localization in the \gls{ssb} frame. Adapted from the LDC Manual. The propagation vector is $\mathbf{\hat{k}}$. The polarization vectors are $\mathbf{\hat{u}}$ and $\mathbf{\hat{v}}$.}
    \label{fig:ssb-frame}
\end{figure}

The \gls{ssb} Cartesian coordinate system is defined by $(\vb{x}, \vb{y}, \vb{z})$, such that $(\vb{x}, \vb{y})$ is the plane of the ecliptic. The \gls{ssb} frame is used to express the \gls{lisa} spacecraft coordinates, as well as the binary sky-localization. We introduce the \gls{ssb} spherical coordinates $(\theta, \phi)$ as illustrated in \cref{fig:ssb-frame}, based on the orthonormal basis vectors $(\vu{e}_r, \vu{e}_\theta, \vu{e}_\phi)$. The source localization is parametrized by the \textit{ecliptic latitude} $\beta = \pi / 2 - \theta$ and the \textit{ecliptic longitude} $\lambda = \phi$. The basis vectors read
\begin{subequations}
\begin{align}
    \vu{e}_r &= (\cos \beta \cos \lambda, \cos \beta \sin \lambda, \sin \beta)
    \qc \\
    \vu{e}_\theta &= (\sin \beta \cos \lambda, \sin \beta \sin \lambda, -\cos \beta)
    \qc \\
    \vu{e}_\phi &= (-\sin \lambda, \cos \lambda, 0)
    \qs
\end{align}
\end{subequations}

The propagation vector is $\vu{k} = -\vu{e}_r$. We define the \textit{polarization vectors} as $\vu{u} = -\vu{e}_\phi$ and $\vu{v} = -\vu{e}_\theta$. This produces a direct orthonormal basis in $(\vu{u}, \vu{v}, \vu{k})$.

\subsection{Projection on the constellation arms}
\label{sec:project}

The deformation induced on link 12, measured on \gls{mosa} 12, is denoted as $H_{12}(t)$. It is computed by projecting the \gls{ssb} \gls{gw} strain on the link unit vector $\vu{n}_{12}$ (computed from the spacecraft positions),
\begin{equation}
\begin{split}
    H_{12}(t) &= h_+^\text{SSB}(t) \times \xi_+(\vu{u}, \vu{v}, \vu{n}_{12})
    \\
    &\qquad + h_\times^\text{SSB}(t) \times \xi_\times(\vu{u}, \vu{v}, \vu{n}_{12}) \qc
\end{split}
\label{eq:projected-strain-ssb}
\end{equation}
where we assume that the link unit vector $\vu{n}_{12}$ is constant during the light travel time. The \textit{antenna pattern functions} are given by
\begin{subequations}
\begin{align}
    \xi_+(\vu{u}, \vu{v}, \vu{n}_{12}) &= \qty(\vu{u} \vdot \vu{n}_{12})^2 - \qty(\vu{v} \vdot \vu{n}_{12})^2 \qc
    \\
    \xi_\times(\vu{u}, \vu{v}, \vu{n}_{12}) &= 2 \qty(\vu{u} \vdot \vu{n}_{12}) \qty(\vu{v} \vdot \vu{n}_{12}) \qs
\end{align}
\end{subequations}

Light emitted by spacecraft 2 at $t_2$ reaches spacecraft 1 at $t_1$. These two times $t_1$ and $t_2$ are related by $H_{12}(\vb{x}, t)$,
\begin{equation}
    t_1 \approx t_2 + \frac{L_{12}}{c} - \frac{1}{2c} \int_0^{L_{12}}{H_{12}(\vb{x}(\lambda), t(\lambda)) \dd{\lambda}} \qs
\label{eq:reception-time-with-H}
\end{equation}
We approximate the wave propagation time to first order as $t(\lambda) \approx t_2 + \lambda/c$. Also, $\vb{x}(\lambda) = \vb{x}_2(t_2) + \lambda \vu{n}_{12}(t_2)$, where $\vb{x}_2(t_2)$ represents the position of the emitter spacecraft at emission time. Using these two expressions, we can further refine $H_{12}$ as
\begin{align}
\begin{split}
    &H_{12} \qty(\vb{x}(\lambda), t(\lambda)) 
    = H_{12} \qty(t(\lambda) - \frac{\vu{k} \vdot \vb{x}(\lambda)}{c})
    \\
    &\qquad = H_{12} \qty(t_2 - \frac{\vu{k} \vdot \vb{x}_2(t_2)}{c} + \frac{1 - \vu{k} \vdot \vu{n}_{12}(t_2)}{c} \lambda) \qs
\end{split}
\label{eq:link-deformation-function-of-x}
\end{align}
Combining \cref{eq:link-deformation-function-of-x,eq:reception-time-with-H} and differentiating the resulting expression with respect to $t_2$ yields the relative frequency shift, $y_{12}$, experienced by light as it travels along link 12,
\begin{equation}
\begin{split}
    y_{12}(t_2) &\approx \frac{1}{2 \qty(1 - \vu{k} \vdot \vu{n}_{12}(t_2))} \Big[ H_{12} \qty(t_2 - \frac{\vu{k} \vdot \vb{x}_2(t_2)}{c})
    \\
    & - H_{12} \qty(t_2 - \frac{\vu{k} \vdot \vb{x}_1(t_1)}{c} + \frac{L_{12}}{c}) \Big]
    \qs
\end{split}
\end{equation}
Here, we have introduced the receiver spacecraft position at reception time $\vb{x}_1(t_1) = \vb{x}_2(t_2) + L_{12} \vu{n}_{12}(t_2)$.
Using $t_1 \approx t_2 + L_{12} / c$ and the fact that the spacecraft moves slowly compared to the propagation timescale, we obtain $\vb{x}_2(t_2) \approx \vb{x}_2(t_1)$ and $\vu{n}_{12}(t_1) \approx \vu{n}_{12}(t_2)$. Finally,
\begin{equation}
\begin{split}
    y_{12}(t_1) &\approx \frac{1}{2 \qty(1 - \vu{k} \vdot \vu{n}_{12}(t_1))} \Big[ H_{12} \Big(t_1 - \frac{L_{12}(t_1)}{c} 
    \\ & - \frac{\vu{k} \vdot \vb{x}_2(t_1)}{c} \Big) - H_{12} \qty(t_1 - \frac{\vu{k} \vdot \vb{x}_1(t_1)}{c}) \Big]
    \qc
\end{split}
\label{eq:instrument-response-to-gw}
\end{equation}
where the equation for $y_{12}$ is now solely a function of reception time $t_1$. Combining \cref{eq:projected-strain-ssb,eq:instrument-response-to-gw} gives $y_{12}$ as a function of $t_1$ in terms of the \gls{gw} strain.

The $y_{ij}$ time series along each of the six links are the final quantities output of the projection step. They are then combined in various ways to compute the \gls{tdi} observables. 

\subsection{Time-delay interferometry}
\label{sec:TDI}

\Gls{tdi} combinations are defined as linear combinations of time-shifted measurements. The first and second-generation Michelson combinations, $X_1$ and $X_2$, are given by~\cite{Tinto:2003vj},
\begin{align}
    \begin{split}
        X_1 &= 
        y_{13} + \delay{13} y_{31} + \delay{131} y_{12} + \delay{1312} y_{21}
        \\
        &\quad - [y_{12} + \delay{12} y_{21} + \delay{121} y_{13} + \delay{1213} y_{31}]
        \qc
    \label{eq:tdi1}
    \end{split}
    \\
    \begin{split}
        X_2 &= X_1 + \delay{13121} y_{12} + \delay{131212} y_{21} + \delay{1312121} y_{13}
        \\
        &\quad+ \delay{13121213} y_{31} - [\delay{12131} y_{13} + \delay{121313} y_{31}
        \\
        &\quad+ \delay{1213131} y_{12} + \delay{12131312} y_{21}]
        \qs
    \label{eq:tdi2}
    \end{split}
\end{align}
Delay operators are defined by
\begin{equation}
    \delay{ij} x(t) = x(t - L_{ij}(t))
    \qc
\end{equation}
where $L_{ij}(t)$ is the delay time along link $ij$ at reception time $t$. Because light travel times evolve slowly with time, we compute chained delays as simple sums of delays rather than nested delays, i.e.,
\begin{equation}
    \delay{i_1, i_2, \dots, i_n} x(t) = x\qty(t - \sum_{k=1}^{n-1}{L_{i_k i_{k+1}}(t)} )
    \qs
\end{equation}
While this approximation cannot be used to study laser-noise suppression upstream of the \gls{lisa} data analysis, it is sufficient when computing the \gls{gw} response function. Note that these equations are left unchanged (up to a sign) by reflection symmetries. However, applying the three rotations generates the three Michelson combinations, $X,Y,Z$, for both generations.

These Michelson combinations have correlated noise properties. An uncorrelated set of \gls{tdi} variables, $A, E, T$, can be obtained from linear combinations of $X,Y,Z$ given by~\cite{Vallisneri2005}
\begin{subequations}
\begin{align}
    A =& \frac{1}{\sqrt{2}}\left(Z-X\right) \qc \\
    E =& \frac{1}{\sqrt{6}}\left(X-2Y+Z\right) \qc \\
    T =&\frac{1}{\sqrt{3}}\left(X+Y+Z\right) \qs
\end{align}
\end{subequations}
The inner products from \cref{eq:like} are really sums over the three channels,
\begin{equation}
    \braket{a}{b} = \sum_{i = A,E,T} \braket{a^i}{b^i}
    \qs
\end{equation}
Note that $A,E,T$ are only exactly orthogonal (or uncorrelated in noise properties) in the equal-armlength limit. In this exploratory work, we maintain the use of $A,E,T$ even in the limit of breathing constellation arms, because we are not considering additive noise and believe this approximation is good enough to inform us of the effect of orbital assumptions on the analysis. In future work, and for the actual \gls{lisa} analysis, we will have to determine the noise information and properly compute the likelihood with off-diagonal terms representing the correlation between the various \gls{tdi} observables~\cite{Vallisneri:2012np}. 

\subsection{Orbital trajectories}
\label{sec:orbitaltraj}

The orbital trajectory of the \gls{lisa} constellation affects the projections of the \glspl{gw} onto the constellation arms, as well as the computation of the \gls{tdi} observables through the armlengths (or, equivalently, the light travel times along each link).

In this study, we focus on two classes of orbits. The standard \textit{equal-armlength} orbital configuration is used in most data analysis codes. The three spacecraft follow heliocentric Keplerian orbits that leave armlengths constant to leading order in the orbit eccentricity~\cite{Dhurandhar2005}. This is realized when the constellation plane has an angle of \SI{60}{\degree} with the ecliptic. In our study, the semi-major axis is set to \SI{1}{\astronomicalunit} and the mean inter-spacecraft distance to the nominal value of \SI{2.5E9}{\meter}. In the frequency domain, which is used for the vast majority of current \gls{lisa} analysis techniques, the armlength is strictly fixed to a single constant value. This is necessary to derive the simpler analytic \gls{tdi} expressions used in frequency-domain template generation~\cite[see, e.g.,][]{Marsat:2018oam}. In the time domain, the ``equal armlengths'' are only nearly constant, but their variations remain much smaller than the typical values seen in realistic orbits (see below).

The other class of orbits we examine are numerically generated orbits provided by the \gls{esa}~\cite{Martens:2021phh}. These orbits take into account all relevant bodies in the Solar system, and are optimized to minimize, amongst others, constellation breathing and fuel consumption necessary to insert the constellation in the correct orbit.

For both equal-armlength and \gls{esa} orbital configurations, the light propagation delay time series along the six constellation links are computed by \texttt{LISA Orbits}~\cite{lisaorbits}, using a second-order post-Minkowskian expansion of the light travel time equation.

One issue in trying to compare equal-armlength orbits to accurate orbits is finding the proper parameterization of the equal-armlength configuration that minimizes any intrinsic bias between the two models. We used the Nelder-Mead optimization algorithm to determine the two free parameters for equal-armlength orbits, chosen as $\lambda_1$, the longitude of the periastron for spacecraft 1, and $m_1$, the mean anomaly at initial time for spacecraft 1. The cost function $f(\lambda_1, m_1)$ was written as the sum, for each spacecraft~$i$, of the absolute distances between the spacecraft positions according to equal-armlength orbits $\vb{x}_i^\text{EQ}$ and their true positions given by accurate \gls{esa} orbits $\vb{x}_i^\text{ESA}$, integrated over four years of mission,
\begin{equation}
    f(\lambda_1, m_1) = \int_{t = 0}^{\SI{4}{yr}}
    \sum_{i=1,2,3}{\norm{\vb{x}_i^\text{EQ}(\lambda_1, m_1, t) - \vb{x}_i^\text{ESA}(t)}}
    \qs
\end{equation}
To check the result, we run the estimator \num{10} times, with different initial guesses for $\lambda_1$ and $m_1$ randomly chosen between \num{0} and $2 \pi$. In all cases, the algorithm converges quickly to the same set of parameters in less than \num{100} iterations.

\begin{figure}
    \centering
    \includegraphics[width=0.9\columnwidth]{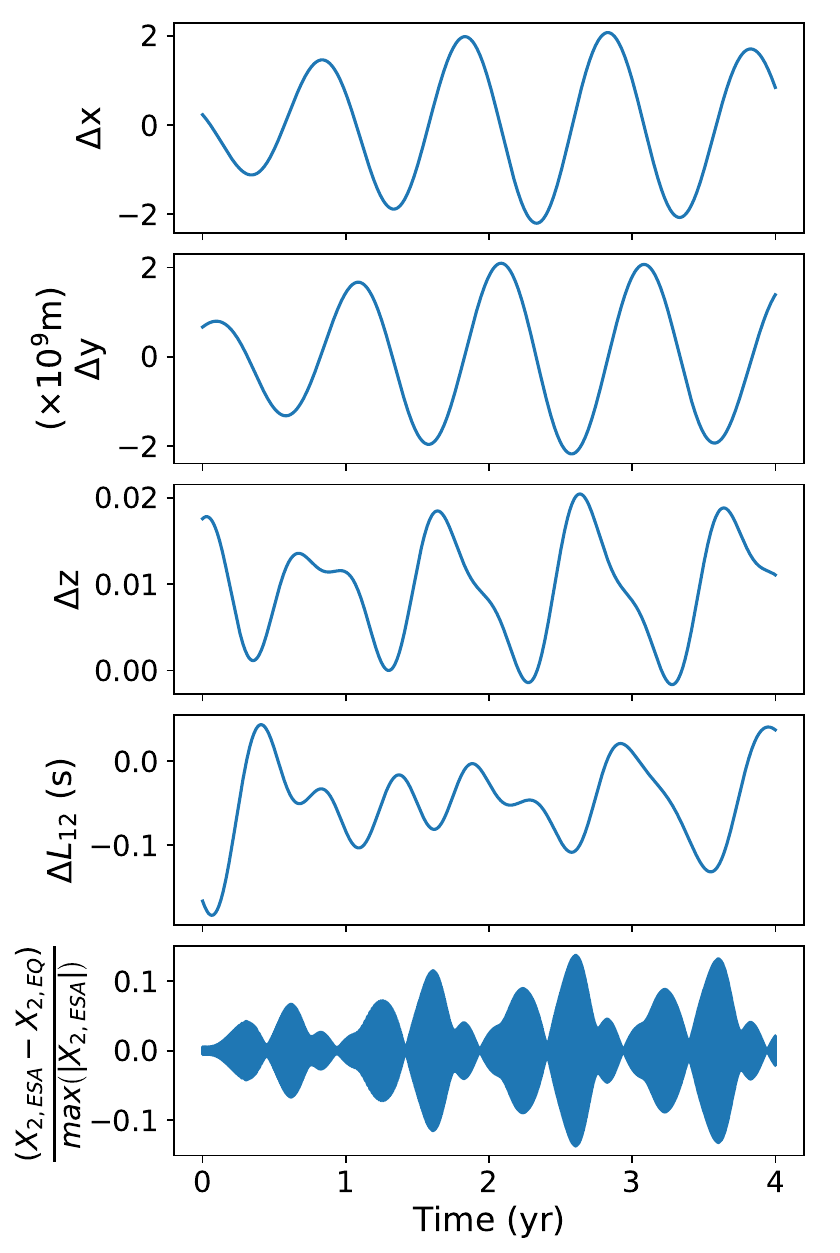}
    \caption{Orbital trajectory difference as determined from the two orbital configurations tested: equal-armlength orbits (``EQ'') and orbits from \gls{esa} (``ESA''). The top three plots show, from top to bottom, the difference in the $x$, $y$, and $z$ coordinates of spacecraft~1's position over four years, in the ecliptic reference frame. The second plot from the bottom shows the difference in the light travel time along link~12. The bottom plot shows an example of relative differences in the \gls{tdi} $X_2$ channel between the two orbital models, over four years of observation, for a binary with parameters $\{ A, f_0, \dot{f}_0, \iota, \phi_0, \psi, \lambda, \beta \}=\{\num{E-22}, \SI{10}{\milli\Hz}, \SI{1.4E-14}{\Hz\per\second}, \num{1.11}, \num{4.58}, \num{0.45}, \num{5.23}, \num{1.23} \}$.}
    \label{fig:orbit_plot}
\end{figure}

Examples of the difference between the two orbital configurations in satellite 1's orbital parameters and light travel time for one link are shown in \cref{fig:orbit_plot}. All three physical coordinates ($x,y,z$) are given in the ecliptic coordinate frame. The maximum and mean difference in $x$ between the two orbits are $\sim \SI{0.04}{\astronomicalunit}$ and $\sim \SI{0.008}{\astronomicalunit}$, respectively. The $y$ coordinate differences have roughly the same properties. The $z$-coordinate maximum difference is $\sim \SI{4E-4}{\astronomicalunit}$ and the mean difference is $\sim \SI{E-4}{\astronomicalunit}$. These spatial coordinate properties are consistent across all three spacecraft. The delay times for link~12 show an expected non-periodic difference between orbits. The delay times for the equal-armlength orbits are roughly constant with a variation range of $\sim \SI{1E-3}{\second}$. The \Gls{esa} orbit delay time for link~12 varies from $\sim \SI{8.154}{\second}$ to $\sim \SI{8.387}{\second}$ (range of $\sim \SI{0.232}{\second}$), clearly displaying a non-periodic behavior. 

The orbit of the \gls{lisa} constellation is included in the $\Lambda$ model parameter from \cref{eq:bayes}. In the case of equal-armlength orbits, $\Lambda$ will be an incorrect model that will incur some mismatch against the true waveforms. A visual comparison of the \gls{tdi} $X_2$ channel between the two models, over the course of observation, is shown in \cref{fig:orbit_plot}. We will address this quantitatively in \cref{sec:posterior}.

\subsection{Implementation}
\label{sec:implement}

Codes performing these generic time-domain calculations have been available for some time~\cite{Vallisneri:2004bn}. Our implementation of this process takes advantage of strong computational acceleration using \glspl{gpu}. Our code, \texttt{fastlisaresponse} can be found on \href{https://github.com/mikekatz04/lisa-on-gpu}{GitHub}~\cite{fastlisaresponse}. All computations of the waveform, response function, and likelihood are performed on \glspl{gpu}, only transferring the final value of the likelihood back to the \acrshort{cpu} processing stream. The simple Galactic binary waveform is produced using \texttt{CuPy}, an effective drop-in replacement for \texttt{NumPy}~\cite{Walt2011} built for NVIDIA \glspl{gpu}.

The response function is coded in \texttt{C++/CUDA} and wrapped into \texttt{Python} with a special \texttt{Cython}~\cite{Cython} setup script from~\cite{CUDAwrapper}. The input orbital information is interpolated and then evaluated at each evenly-spaced time step in the observed data stream. This information is pre-computed and stored for the entirety of the sampling run. All time points in all six projections are evaluated in parallel on a separate thread of the \gls{gpu}. The projections require interpolation of $h_+^\text{SSB}$ and $h_\times^\text{SSB}$ at desired time points. This is performed using centered Lagrange interpolating polynomials of user-defined order. Typically, an order of 25 is used to be conservative. Lower orders, even as low as less than 10, are generally accurate enough and can improve the speed of the computation.

To perform this interpolation, a set of points from the original waveform is required. These points span before and after the desired interpolation time. Typically, this is not a concern for \gls{cpu} programming because efficient memory access is not a limiting factor. For \gls{gpu} programming, with slower memory access from global memory, gathering separate sub-arrays from the original waveform for each point in time where interpolation is needed can be the bottleneck. Therefore, the \gls{gpu} shared memory is leveraged by storing sub-arrays spanning enough length for use by multiple time points during the computation. The use of shared memory led to a crucial improvement in the overall performance of this code.

The same process is performed to compute the \gls{tdi} variables, by running all time points for all \gls{tdi} delay combinations (\cref{eq:tdi1,eq:tdi2}) in parallel along separate threads. The same Lagrange polynomials are used to interpolate the six projections to form the proper \gls{tdi} combinations.

To check the correctness of \texttt{fastlisaresponse}, we compared, for a small set of Galactic binary strain as input, the resulting \gls{tdi} time series to those obtained on \gls{cpu} with the time-domain generic response code \texttt{LISA GW Response}~\cite{lisagwresponse} and \texttt{PyTDI}~\cite{pytdi}. In all cases we tested, time series matched down to numerical precision after correcting for the different conventions on initial times implemented in these codes.

\begin{figure}
    \centering
    \includegraphics[width=0.9\columnwidth]{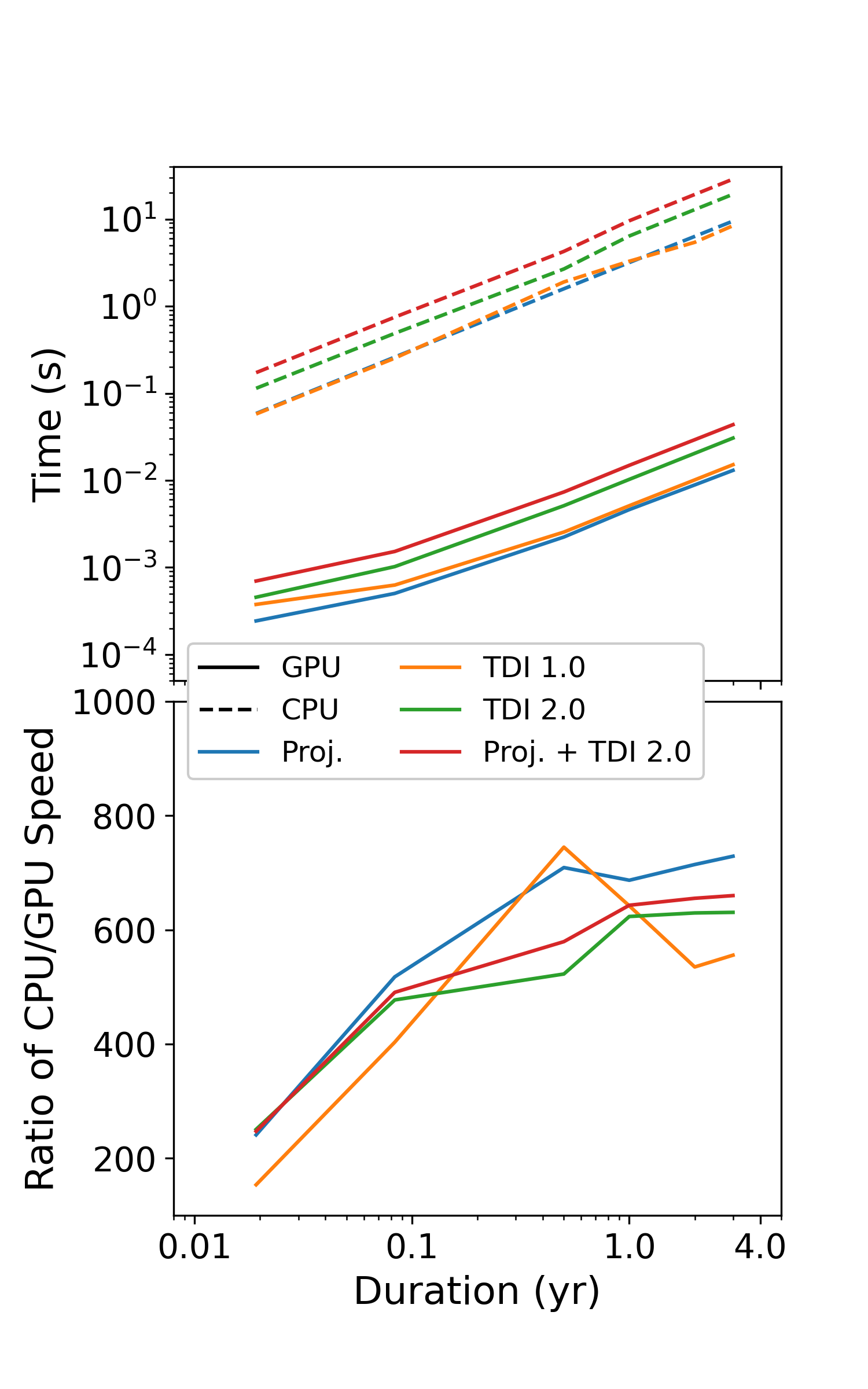}
    \caption{Performance of \texttt{fastlisaresponse}. The horizontal axis represents the simulation duration. In the top panel, \gls{gpu} and \gls{cpu} performance are shown in solid and dashed lines, respectively. The timing of the projection portion is shown in blue. \Gls{tdi} generations 1 and 2 are shown in orange and green, respectively. The total evaluation time of the response, combining the projection portion and \gls{tdi} 2, is shown in red. The bottom panel shows the runtime ratio. These timing tests were performed on an NVIDIA Tesla A100 \gls{gpu} and a single AMD EPYC 7713 \gls{cpu}.}
    \label{fig:speed}
\end{figure}

The speed of this new code is illustrated in \cref{fig:speed} comparing \gls{cpu} to \gls{gpu} performance as a function of the duration of a template waveform. The speed comparison is shown for the projection piece (``Proj.''), \gls{tdi} 1, \gls{tdi} 2, and the full response function combining the projection with \gls{tdi} 2. These tests were performed on a single AMD EPYC 7713 \gls{cpu} processor and an NVIDIA Tesla A100 \gls{gpu}. As the duration of the data stream increases, the \gls{gpu} performance increase levels out at $\sim 600 \times$ faster than the \gls{cpu}. The response function for one year of observation takes $\sim \SI{10}{\milli\second}$ ($\sim \SI{E4}{\milli\second}$ for the \gls{cpu}). The Lagrangian interpolation order does affect this time by a factor of order unity between the minimum and maximum orders tested of 1 and 25, respectively. In current frequency-domain response methods, the response function is usually evaluated in $\sim \si{\milli\second}$ on a \gls{cpu} and $\sim \si{\micro\second}$ when evaluated in batches on a \gls{gpu}~\cite{Katz:2021uax}. The main reason for this efficient speed is the ability to operate with a much sparser or shorter array of frequencies compared to the number of time points in the time-domain signal. Frequency-domain response functions are usually evaluated with $\sim 2^{10}$ points or less compared to the $\sim 2^{22}$ points in the equivalent one-year time-domain evaluation.

\section{Posterior analysis}
\label{sec:posterior}

We analyze the effect of using different orbits within \gls{mcmc} by taking advantage of the computational efficiency of the \gls{gpu} implementation. We tested binaries across a multidimensional grid in \gls{snr}, ecliptic latitude, frequency, and frequency derivative. Two \gls{snr} values were used: an \gls{snr} of 30 was chosen to examine if orbital effects are visible in common, quieter Galactic binary signals. An \gls{snr} of 500 was chosen to represent louder sources, and allow us to ensure that the response function remains accurate enough for our ``best'' sources~\cite{Cornish:2017vip}. Two ecliptic latitudes were examined; one configuration is near-planar to the ecliptic with $\beta = \SI{0.09}{\radian}$; the other is near-polar with $\beta = \SI{1.2}{\radian}$.

In order to span the range of interest of Galactic binary frequencies and frequency derivatives, we tested three independent parameter configurations with $(f, \dot{f})$ $\in$ $\{ (\num{7E-4}, \num{1E-19})$, $(\num{2E-3}, \num{1.2E-17})$, $(\num{1E-2}, \num{1.4E-14}) \}$ (frequency in \si{\Hz} and frequency derivative in \si{\Hz\per\second}). The three configurations were chosen to represent a binary at low-frequency and low-frequency derivative (chirp mass of $\sim~\SI{0.08}{\solarmass}$); one binary at middle frequency and middle frequency derivative (chirp mass of $\sim~\SI{0.2}{\solarmass}$); and one binary at high frequency and frequency derivative (chirp mass of $\sim~\SI{0.75}{\solarmass}$). The amplitude for each source tested was adjusted to achieve the desired \gls{snr} value.

The other parameters were randomly chosen from their respective prior distributions (see below). Our tests were computed with $\{ \iota, \phi_0, \psi, \lambda \}=\{ \num{1.11}, \num{4.58}, \num{0.45}, \num{5.23} \}$ (in \si{\radian}). Prior to performing \gls{mcmc}, we tested a larger grid of values with basic mismatch calculations. It was clear from these computations that the main factors affecting the mismatch were the frequency, frequency derivative, and the ecliptic latitude. Therefore, we believe our grid is representative of the parameter space for Galactic binaries.

For each point in the grid, we performed three \gls{mcmc} runs. For all three runs, a signal built from \gls{esa} orbits was injected. Then, each run can test one of three different orbital configurations in the template waveforms. The first template was generated with \gls{esa} orbits. This run acts as a \textit{control} with a template-signal fitting factor of 1, indicating no inherent bias in the template production. The second run involved using the equal-armlength orbits (``EQ'') to generate the template. In this case, the fitting factor is strictly less than 1, allowing us to investigate the parameter-estimation bias incurred from using incorrect \gls{lisa} constellation orbital properties. A third test that we performed is a middle ground between the first two, where the projection operation (\cref{sec:project}) is performed with \gls{esa} orbits, while the \gls{tdi} operation (\cref{sec:TDI}) is computed with equal-armlength orbits. This model tests whether switching to accurate orbits in the projection portion of the response is sufficient for proper data analysis (this would allow for accurate projections while maintaining the efficiency of analytic \gls{tdi} methods that require an equal-armlength configuration). We will refer to this \textit{hybrid} orbital arrangement as ``ESAEQ''.

The prior distributions used on the parameters are given in \cref{tb:priorinfo}. Uniform distributions were used for parameters $f_0$, $\phi_0$, $\psi$, and $\lambda$. The inclination ($\iota$) prior is uniformly distributed in the cosine of the inclination angle. The ecliptic latitude ($\beta$) prior is uniform in $\sin{\beta}$. The prior on $A$ is log-uniform between $A^*\times10^{-2}$ and $A^*\times10^{2}$, where $A^*$ is the injected amplitude. Like the prior on $A$, the prior on the frequency derivative ($\dot{f}_0$) is adjusted based on the injection to ensure physically reasonable, but also encapsulating, limits on the chirp mass of the system. The frequency derivative is given a uniform prior, with boundaries computed as $\dot{f}_0^{\mathrm{min},\mathrm{max}} = \dot{f}_\mathrm{GW}(f_0, \mathcal{M}_c^{\mathrm{min},\mathrm{max}})$, see \cref{eq:fdot}, with physically representative $\mathcal{M}_c^\mathrm{min}=10^{-3} M_\odot$ and $\mathcal{M}_c^\mathrm{max}=1 M_\odot$.

\begin{table}
\centering
\begin{tabular}{@{}ccc@{}}
\toprule
Parameter & Lower Bound & Upper Bound \\
\midrule
$\ln{A}$ & $\ln(A^*\times\num{E-2})$ & $\ln(A^*\times\num{E+2}) $ \\
$f_0$ (\si{\milli\hertz}) & 0.5 & 12 \\
$\dot{f}_0$ (\si{\hertz^2}) & $\dot{f}(f_0^*, \mathcal{M}_c=\num{E-3}M_\odot)$ & $\dot{f}(f_0^*, \mathcal{M}_c=\num{E0}M_\odot)$ \\
$\cos{\iota}$ & $-1$ & 1 \\
$\phi_0$ & 0 & $2\pi$ \\
$\psi$ & 0 & $\pi$ \\
$\lambda$ & 0 & $2\pi$ \\
$\sin{\beta}$ & $-1$ & 1 \\
\bottomrule
\end{tabular}
\caption{Prior distributions used in our analysis. The priors on $f_0$, $\phi_0$, $\psi$, and $\lambda$ are uniform distributions. The prior on the amplitude, $A$, is log-uniform, spanning two orders of magnitude above and below the injected value of $A^*$. The inclination prior ($\iota$) and the ecliptic latitude prior ($\beta$) are uniform in $\cos{\iota}$ and $\sin{\beta}$, respectively. The prior on $\dot{f}_0$ is adjusted to ensure reasonable values for the chirp of a Galactic binary, given an initial frequency $f_0$.}
\label{tb:priorinfo}
\end{table}

\begin{figure}
    \centering
      \includegraphics[width=\columnwidth]{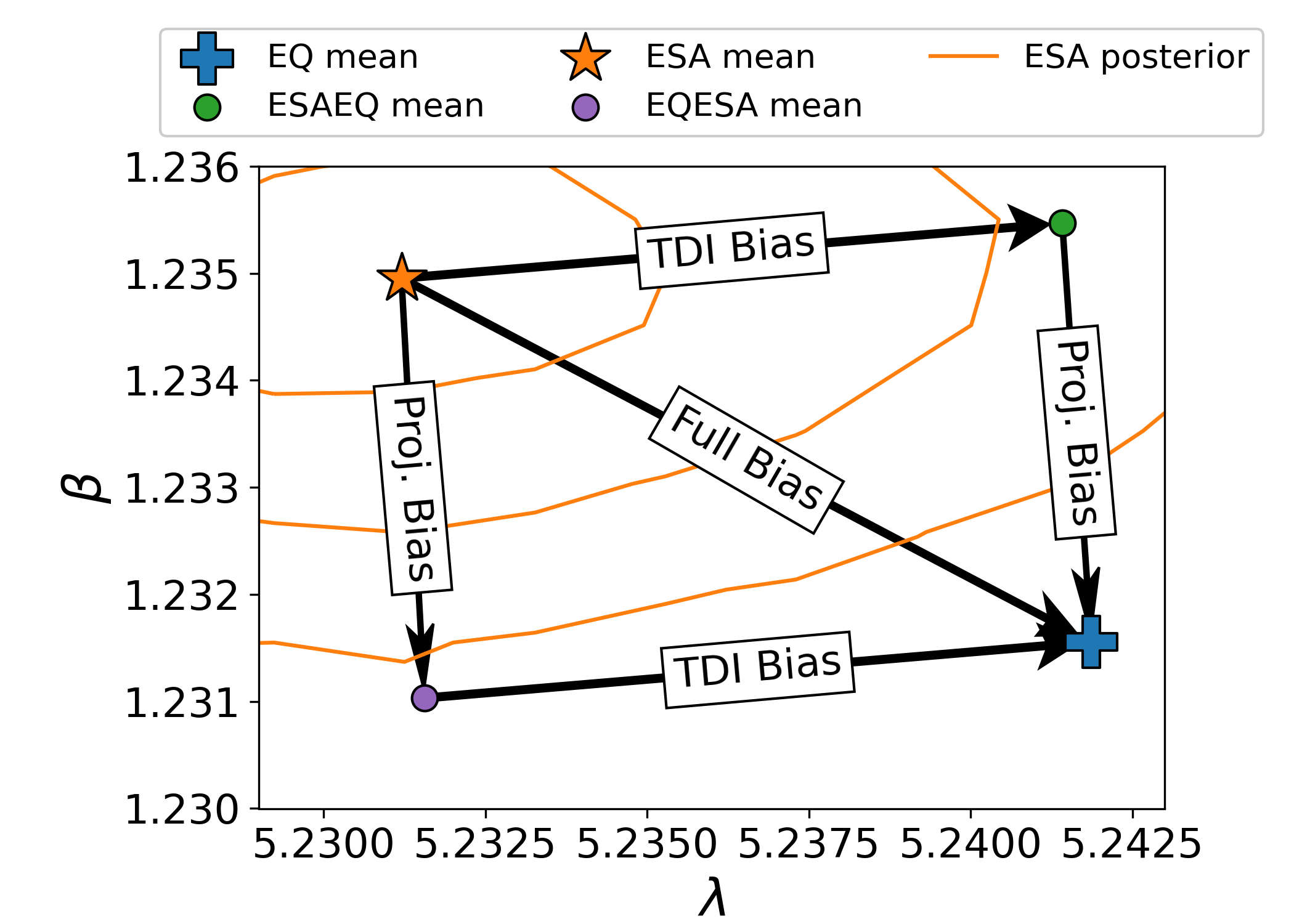}
      \caption{An example of bias created using inaccurate orbital information. The true sky localization is represented by the orange star, coincidental with the mean of the ESA template run. The orange lines represent the $1\sigma$, $2\sigma$, and $3\sigma$ contours for the two-dimensional marginalized distribution using accurate ESA orbits. The EQ template mean is shown with a blue plus. Connecting the ESA mean to the EQ mean yields the overall bias when using the EQ orbital configuration. ESAEQ and EQESA configuration means are shown with green and purple dots, respectively. Connecting ESA to ESAEQ represents the bias incurred by using equal-armlength orbits in the projection portion of the response function. Alternatively, connecting the ESA mean to the ESAEQ mean reveals the bias when using equal-armlength orbits in the \gls{tdi} portion of the response function. The two bias components seem, visually, to add linearly. This is because the overall bias is small enough. In general, the overall bias from using incorrect orbits is inherently non-linear, as the \gls{tdi} portion operates directly on the projections.}
    \label{fig:biasbreakdown}
\end{figure}

The assessment of bias from \gls{mcmc} results is visualized in \cref{fig:biasbreakdown}. In this figure, we examine a Galactic binary with parameters $\{ \text{SNR}, \beta, (f, \dot{f}) \} = \{ \num{500}, \num{1.2}, (\SI{7E-4}{\Hz}, \SI{1E-19}{\Hz\per\second}) \}$. We focus the estimation of ecliptic longitude and latitude, shown on the horizontal and vertical axes, respectively. The mean estimates from parameter estimation runs with four different orbital configurations are shown with scatter points. The four configurations are ESA (orange star), EQ (blue plus), ESAEQ (green dot), and EQESA (purple dot). As the name implies, EQESA involves using equal-armlength orbits for the projections and \gls{esa} orbits for the \gls{tdi} computation.

Connecting the ESA mean to the EQ mean in \cref{fig:biasbreakdown} gives the overall bias on the sky location when using equal-armlength orbits. The contributions of the projection and \gls{tdi} portions of the response to the overall bias can also be determined: the projection bias connects the mean of the ESA configuration to the mean of the EQESA configuration. The arrow from the ESA mean to the ESAEQ mean gives the bias from \gls{tdi}. While the visualization appears here to indicate that the biases add linearly, this process is inherently non-linear as any projection bias will feed into the \gls{tdi} computation. In addition, the two-dimensional marginalized posterior distribution for the ESA orbital configuration is shown as its $1\sigma$, $2\sigma$, and $3\sigma$ contours (orange lines). For the source studied here, the bias in the ESAEQ configuration is beyond $1\sigma$, indicating the \gls{tdi} bias from equal-armlength orbits is significant for higher \gls{snr} sources. We will discuss these effects further below.

\begin{figure*}
    \centering
    \begin{tabular}{cc}
    \includegraphics[width=\columnwidth]{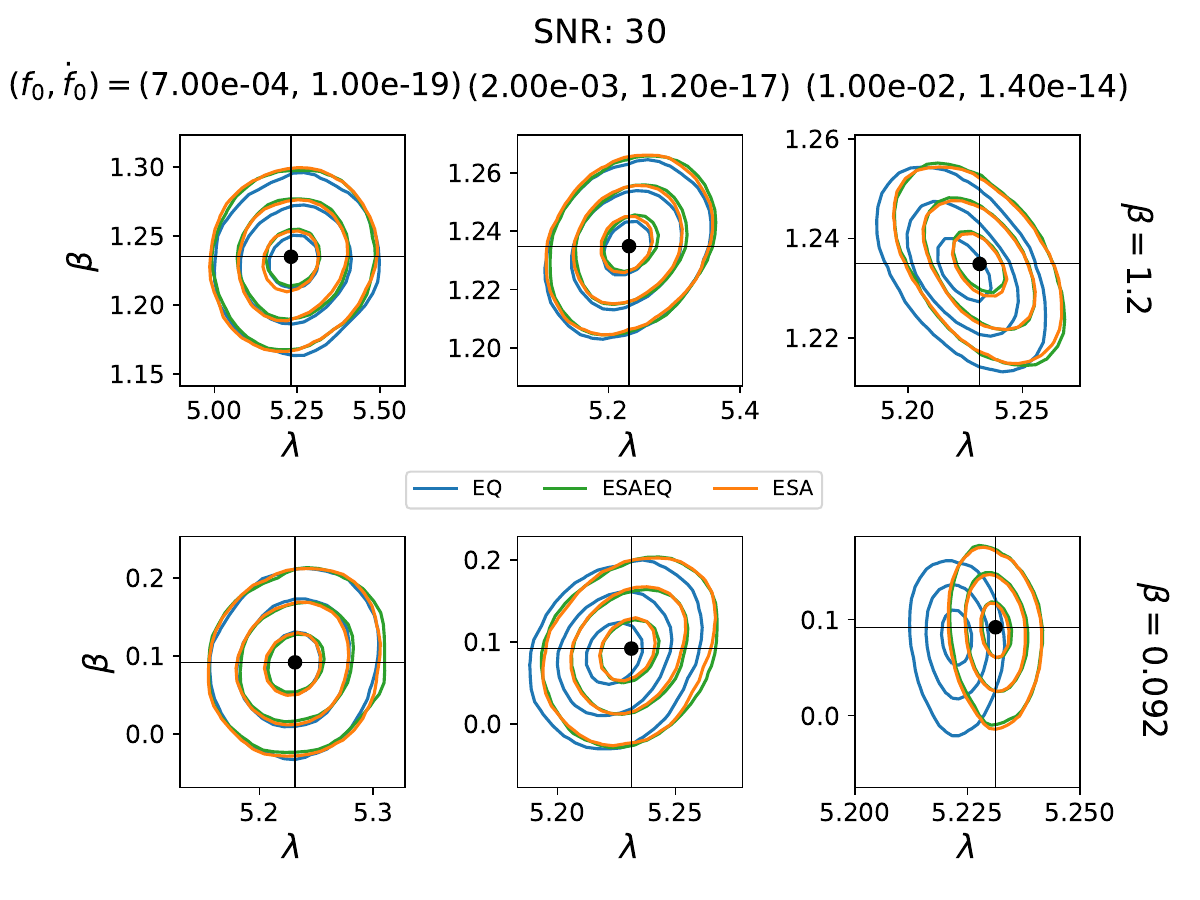} &
    \includegraphics[width=\columnwidth]{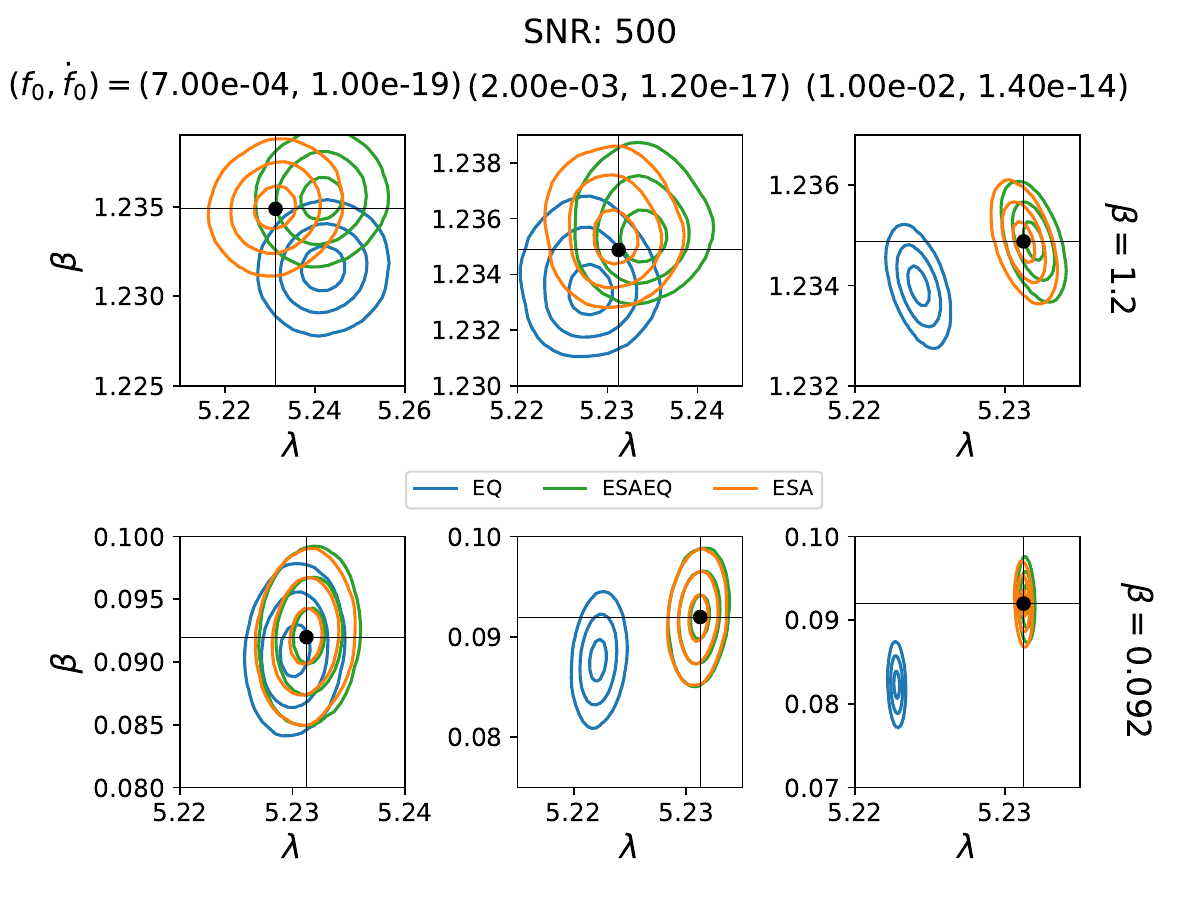}
    \end{tabular}
    \caption{Two-dimensional marginalized posterior distributions on the sky location, across a variety of injected parameters and \glspl{snr}. The left set of plots shows an \gls{snr} of 30 and the right set an \gls{snr} of 500. The lower \gls{snr} is chosen to resemble a high number of detectable and characterizable Galactic binaries. The higher \gls{snr} represents the best sources we expect to detect. Within each \gls{snr} set, the columns represent $(f_0,\dot{f}_0)$ combinations of $(\num{7E-4}, \num{1E-19})$, $(\num{2E-3}, \num{1.2E-17})$, $(\num{1E-2}, \num{1.4E-14})$ from left to right, respectively. The top (bottom) row in each \gls{snr} set is for the off-planar (on-planar) sources with $\beta=1.2$ ($\beta=0.092$). Within each subplot, the posterior distributions are shown for three orbital models: ESA in orange, EQ in blue, and ESAEQ in green. The deviation of the mean of the EQ or ESAEQ models compared to the mean of the ESA model gives the bias associated with the two inaccurate orbital configurations.}
    \label{fig:skycomp}
\end{figure*}

\begin{figure*}
    \centering
    \begin{tabular}{cc}
    \includegraphics[width=\columnwidth]{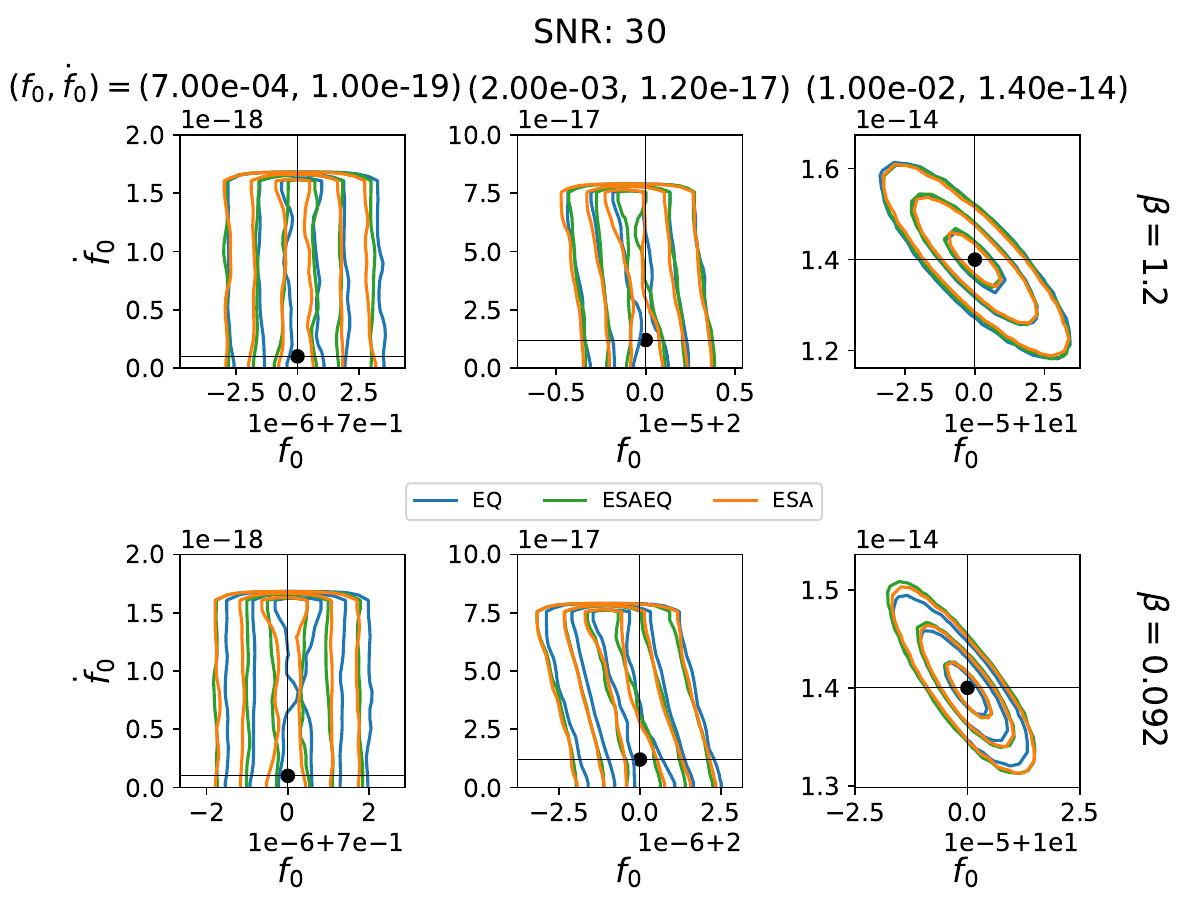} &
    \includegraphics[width=\columnwidth]{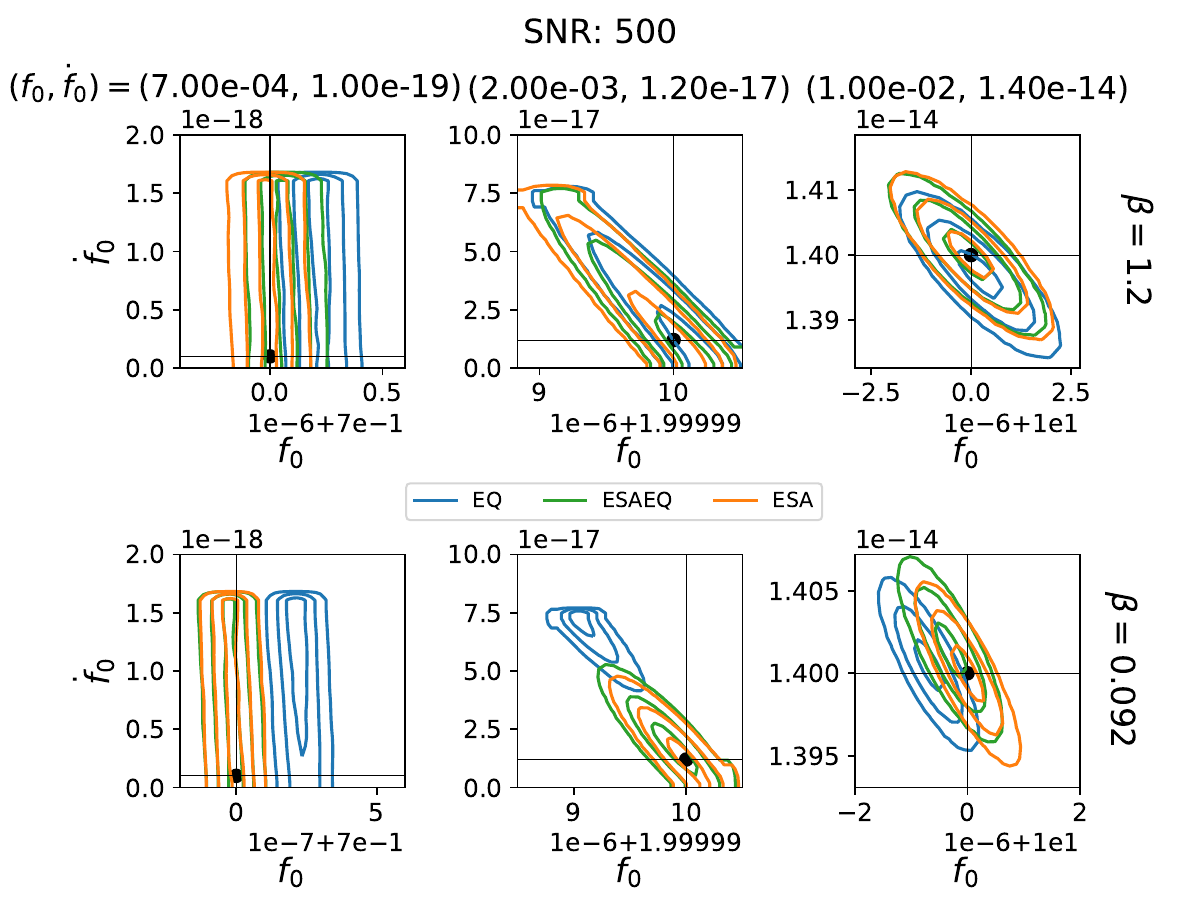}
    \end{tabular}
    \caption{Two-dimensional marginalized posterior distributions, with $f_0$ along the horizontal axis and $\dot{f}_0$ along the vertical axis. These posteriors are shown across a variety of injected parameters and \gls{snr} combinations. C.f. \cref{fig:skycomp} for details on the arrangement of the subplots.}
    \label{fig:ffdotcomp}
\end{figure*}

This bias examination over the full posterior distributions was performed for each point on our multidimensional parameter grid. Here, we highlight important two-dimensional marginalized posterior distributions to help illustrate the effect of using incorrect orbits. \Cref{fig:skycomp,fig:ffdotcomp} show the two-dimensional posteriors in $f_0-\dot{f}_0$ (initial frequency and frequency derivative) and $\beta-\lambda$ (sky localization angles), respectively. These parameter are of high scientific interest, as they characterize the intrinsic Galactic binary systems and provide opportunities for electromagnetic counterpart observations.

These figures are grouped by \gls{snr} (\num{30} on the left and \num{500} on the right). Within each \gls{snr} group, $\beta = \num{1.23}$ is shown in the first row and $\beta = \num{0.09}$ is shown in the bottom row. Then, from left to right, within each row, are the low $f_0-\dot{f}_0$, middle $f_0-\dot{f}_0$, and high $f_0-\dot{f}_0$ cases. For each \gls{mcmc} run shown in these plots, the log-likelihood and mismatch marginalized over phase for the misaligned orbits at the injection parameters are given in \cref{tb:llmm}. Also in \cref{tb:llmm}, the log-likelihood and mismatch are given at the best-fit location (maximum log-likelihood) found by the sampler.

\begin{center}
\begin{table*}
	\begin{tabularx}{0.8\textwidth}{
	| >{\hsize=.5\hsize}>{\centering\arraybackslash}X
	| >{\hsize=.5\hsize}>{\centering\arraybackslash}X 
	| >{\centering\arraybackslash}X 
	| >{\centering\arraybackslash}X 
	| >{\centering\arraybackslash}X 
	| >{\centering\arraybackslash}X 
	| >{\centering\arraybackslash}X 
	| >{\centering\arraybackslash}X|
	}
\hline
\gls{snr} & $\beta$ & $f_0-\dot{f}_0$ & Orbit & Inj. $\ln{\mathcal{L}}$ & Inj. MM & Best $\ln{\mathcal{L}}$ & Best MM \\ 
\hline
\hline30 & 1.2 & low & EQ & \num{-2.9E-01} & \num{1.7E-02} & \num{-2.0E-01} & \num{-1.3E-03} \\ 
\hline 
30 & 1.2 & low & ESAEQ & \num{-8.0E-02} & \num{1.1E-02} & \num{-1.9E-01} & \num{-1.3E-02} \\ 
\hline 
30 & 1.2 & mid & EQ & \num{-3.2E-01} & \num{1.7E-02} & \num{-2.3E-01} & \num{5.2E-03} \\ 
\hline 
30 & 1.2 & mid & ESAEQ & \num{-7.9E-02} & \num{1.1E-02} & \num{-1.9E-01} & \num{1.8E-03} \\ 
\hline 
30 & 1.2 & high & EQ & \num{-1.1E+00} & \num{1.4E-02} & \num{-2.7E-01} & \num{-7.0E-03} \\ 
\hline 
30 & 1.2 & high & ESAEQ & \num{-7.0E-02} & \num{7.8E-03} & \num{-1.6E-01} & \num{3.2E-03} \\ 
\hline 
30 & 0.092 & low & EQ & \num{-3.9E-01} & \num{1.4E-02} & \num{-2.8E-01} & \num{-2.3E-03} \\ 
\hline 
30 & 0.092 & low & ESAEQ & \num{-9.8E-02} & \num{1.0E-02} & \num{-1.9E-01} & \num{-7.1E-04} \\ 
\hline 
30 & 0.092 & mid & EQ & \num{-6.8E-01} & \num{1.4E-02} & \num{-3.1E-01} & \num{7.4E-03} \\ 
\hline 
30 & 0.092 & mid & ESAEQ & \num{-9.7E-02} & \num{1.0E-02} & \num{-1.5E-01} & \num{-1.2E-02} \\ 
\hline 
30 & 0.092 & high & EQ & \num{-8.2E+00} & \num{2.0E-02} & \num{-3.8E-01} & \num{-1.2E-02} \\ 
\hline 
30 & 0.092 & high & ESAEQ & \num{-8.6E-02} & \num{7.7E-03} & \num{-2.1E-01} & \num{5.9E-03} \\ 
\hline 
500 & 1.2 & low & EQ & \num{-8.0E+01} & \num{1.7E-02} & \num{-2.8E+01} & \num{2.5E-04} \\ 
\hline 
500 & 1.2 & low & ESAEQ & \num{-2.2E+01} & \num{1.1E-02} & \num{-4.5E+00} & \num{2.9E-04} \\ 
\hline 
500 & 1.2 & mid & EQ & \num{-8.8E+01} & \num{1.7E-02} & \num{-3.3E+01} & \num{-9.8E-05} \\ 
\hline 
500 & 1.2 & mid & ESAEQ & \num{-2.2E+01} & \num{1.1E-02} & \num{-6.3E+00} & \num{-3.3E-04} \\ 
\hline 
500 & 1.2 & high & EQ & \num{-3.1E+02} & \num{1.4E-02} & \num{-3.0E+01} & \num{4.8E-04} \\ 
\hline 
500 & 1.2 & high & ESAEQ & \num{-1.9E+01} & \num{7.8E-03} & \num{-4.3E+00} & \num{9.4E-05} \\ 
\hline 
500 & 0.092 & low & EQ & \num{-1.1E+02} & \num{1.4E-02} & \num{-5.9E+01} & \num{-6.3E-04} \\ 
\hline 
500 & 0.092 & low & ESAEQ & \num{-2.7E+01} & \num{1.0E-02} & \num{-1.3E+01} & \num{5.2E-04} \\ 
\hline 
500 & 0.092 & mid & EQ & \num{-1.9E+02} & \num{1.4E-02} & \num{-5.3E+01} & \num{3.2E-04} \\ 
\hline 
500 & 0.092 & mid & ESAEQ & \num{-2.7E+01} & \num{1.0E-02} & \num{-1.3E+01} & \num{7.7E-04} \\ 
\hline 
500 & 0.092 & high & EQ & \num{-2.3E+03} & \num{2.0E-02} & \num{-6.5E+01} & \num{2.2E-04} \\ 
\hline 
500 & 0.092 & high & ESAEQ & \num{-2.4E+01} & \num{7.7E-03} & \num{-8.9E+00} & \num{-4.2E-04} \\ 
\hline 
	\end{tabularx}
\caption{Log-likelihood and mismatch information for all parameter-estimation runs tested. The first four columns label each run with the \gls{snr}, ecliptic latitude ($\beta$), $f_0-\dot{f}_0$ regime, and orbital configuration from left to right, respectively. The next two columns labelled ``Inj. $\ln{\mathcal{L}}$'' and ``Inj. MM'' give the log-likelihood and mismatch at the injection parameters marginalized over the initial phase. The final two columns represent the log-likelihood and mismatch at the best-fit location found by the sampler (not marginalized over phase). A mismatch of less than zero (overlap > 1) can occur when the fitting-factor of the model is not 1 as in the cases displayed here.}\label{tb:llmm}
\end{table*}
\end{center}

The resulting observations are fairly consistent across all three figures. Low-\gls{snr} sources generally show minimal relative bias from orbital misalignment. This does not strictly hold for the sky localization distributions of the high and middle $f_0-\dot{f}_0$ sources. For the high $f_0-\dot{f}_0$ source at higher polar latitude, this effect barely reaches $1\sigma$ deviation for the EQ comparison, and is almost entirely in the ecliptic longitude. For the nearly-planar source localization, the relative bias reaches beyond $2\sigma$. This is expected, as a larger projection of the \gls{gw} signal along the plane of \gls{lisa} motion helps constraining more tightly the ecliptic longitude. In the middle $f_0-\dot{f}_0$ configurations, the near-planar latitude system displays a bias that is close to, but slightly less than, $1\sigma$. The nearly-polar case shows minimal bias. The frequency dependence appears clearly: tending towards higher frequencies at lower \gls{snr} will produce observable biases in the source localization. In addition to these findings on the EQ configuration, we observe that the ESAEQ configuration is perfectly acceptable for lower \gls{snr}, as it does not produce any observable bias. 

In all high-\gls{snr} cases, the EQ templates produce strong biases on the sky localization of the sources with higher frequencies. The ESAEQ recovery of the signal is biased by $1\sigma$ or greater for all frequency configurations at high ecliptic latitudes; this reveals that the \gls{tdi} portion alone contributes an observable bias on these parameters. At lower ecliptic latitudes, in the ESAEQ configuration, smaller the bias (less than $1\sigma$) are observed for middle and low $f_0-\dot{f}_0$. For the high $f_0-\dot{f}_0$ case at near-planar ecliptic latitude, the bias in the ESAEQ case does reach $1\sigma$.

These high-\gls{snr} cases also produce observable biases on the other parameters (there are minimal biases on the other parameters at low \gls{snr}). As an example, \cref{fig:ffdotcomp} shows posterior distributions for $f_0$ and $\dot{f}_0$. In the low $f_0-\dot{f}_0$ cases, $\dot{f}_0$ is not well-constrained such that we only study the effect of orbital misalignment on $f_0$ distributions. At both latitudes, frequency is biased by more than $3\sigma$ for the EQ configuration. The ESAEQ configuration does not exhibit large biases at low latitudes; at high ecliptic latitude, however, $f_0$ is biased by almost $2 \sigma$. The high $f_0-\dot{f}_0$ posteriors are highly Gaussian, an indication that $\dot{f}_0$ is well constrained. The EQ biases at high latitude are mainly in $\dot{f}_0$ (around $1 \sigma$) with only a small shift in $f_0$ (less than $1 \sigma$). The ESAEQ case does not show any observable bias, indicating that the overall bias is mainly due to the projection portion of the response. For the low-latitude case, we observe an EQ bias almost entirely in frequency ($\sim 3\sigma$), with a noticeable contribution from the ESAEQ setup ($\sim1\sigma$ bias). As expected, the frequency derivatives of high-\gls{snr}, middle $f_0-\dot{f}_0$ sources are only slightly constrained, and highly correlated with $f_0$ (unlike in the low $f_0-\dot{f}_0$ case). At high latitude, the EQ and ESAEQ setups are only biased by about $1\sigma$ or less. The low-latitude case shows that using EQ templates yields very strong biases in frequency and frequency derivative (towards the maximum values allowed by the chosen prior). The ESAEQ template only shows a small bias, indicating that the overall bias originates in the projection portion of the response.

In addition to the parameters discussed above, we also observe a bias on the amplitude $A$ of the source. This is expected, because a slight change in orbital configurations (even while maintaining other parameters fixed) creates slight deviations in the template \gls{snr}; the amplitude space is then explored by the sampler so that the template \gls{snr} matches that of the injected signal, which ultimately leads to bias on the amplitude. The chirp mass can be determined using the posterior distributions for $f_0$ and $\dot{f}_0$, c.f., \cref{eq:fdot}. Then, the chirp mass, $f_0$, and $A$ can be combined to determine the luminosity distance. As a consequence, slight errors on the amplitude (due to variations of the \gls{snr} because of wrong orbital assumptions) will lead to errors on luminosity distance of the source.

\section{Discussion}
\label{sec:discuss}

The creation of a fast generic time-domain response function for \gls{lisa} will finally allow the \gls{lisa} Community to perform full Bayesian inference with accurate \gls{lisa} constellation orbital information. Prior to the creation of this fast response function, all \gls{lisa}-based analyses were either too slow to scale to the level necessary for \gls{mcmc}-style techniques or were performed with templates built from approximated methods. Now, with the construction of any time-domain template that is fast enough for \glspl{mcmc}, there is an immediately implementable response function ready for the task.

This not only produces a more accurate analysis, but it also allows us to perform a wider range of tests to understand the impact of various response approximations. This is an important question, because, as previously mentioned, all current analysis codes use such approximations, such as the equal-armlength setup examined in this paper. The main reason for this is that these orbits are analytic, which allows for a variety of flexible implementations. Further quantifying these systematic effects will help the \gls{lisa} community to understand where its currently available and most efficient tools can be used, as well as how to effectively allocate resources to make necessary improvements to different areas of LISA data analysis.

The advantage of analytic orbits manifests differently in the link projection and \gls{tdi} portions of the \gls{lisa} response. In the projections, due to \gls{lisa}'s time-dependent motion, the positions of the spacecraft have to be computed for a given array of times. In fast models, where the response can be computed on a \textit{sparse grid} in the time-domain (such as \texttt{FastGB}~\cite{Cornish:2007if}), the projections remain tractable with more accurate orbits, because they only involve computing spacecraft locations at given times. When evaluating frequency-domain waveforms and response functions (such as \texttt{bbhx}~\cite{Katz:2021uax}), the time-frequency correspondence of each harmonic mode is used to determine the position of the \gls{lisa} constellation at a given frequency. Therefore, if the orbital information is input into the response code as an interpolant, the projections in the frequency-domain waveforms will also be quite flexible to orbital adjustments. 

During \gls{tdi}, the frequency domain is greatly affected by the equal-armlength approximation. Currently, all \gls{mcmc}-capable response codes evaluate the \gls{tdi} combinations in the frequency domain. Working in the frequency domain and assuming equal-armlengths vastly simplifies the \gls{tdi} computation. In the time-domain response function used in this work, each term in \cref{eq:tdi1} or \cref{eq:tdi2} is interpolated and computed separately and then added together. The frequency-domain equal-armlength response assumes the delays are all constant and equivalent to $L/c$, where $L$ is the chosen rigid armlength of the \gls{lisa} constellation. This reduces the necessary interpolation in the time-domain to basic phase shifts in the frequency-domain, greatly simplifying and accelerating the computation. 

It is important to understand how well the equal-armlength approximation performs in order to understand how useful the current frequency-domain response functions will be to \gls{lisa} analysis. As was presented in the previous section, it is clear equal-armlength orbits will not be accurate enough for parameter estimation purposes. They are mostly fine at lower \glspl{snr}, but at any higher \glspl{snr}, the biases, relative to the increased precision in our parameter estimation, are not acceptable. However, all response codes in use could easily be updated for accurate orbits in the projection portion of the computation. Therefore, if the ESAEQ templates exhibited an acceptably small bias, the frequency-domain response codes would probably remain sufficient for parameter estimation.

However, it is clear that, at higher (but reasonable) \glspl{snr}, even this type of half-approximation does not hold for all sources tested. \textit{This incursion of bias from equal-armlength orbits indicates frequency-domain response codes will have to be updated to a more accurate response formulation for use in parameter estimation}. One potential solution is to perform the analysis in the time-frequency or wavelet domain~\cite{Cornish:2020odn}, but this requires further investigation.

Due to the ability of the EQ templates to produce a high empirical \gls{snr} and reasonable, albeit biased, posteriors, the equal-armlength approximation will probably be useful in search, especially if it remains a much faster computation. This will require verification in future work, especially for sources that require more than one year of observation for detection (our tests in this work are over one year). From \cref{fig:orbit_plot}, we see that the difference in the orbits, as well as the difference in the waveforms that manifest from those orbits, gradually increases over longer durations. 

In this work, we have only tested Galactic binaries, the simplest of \gls{lisa} sources in terms of their waveform complexity. The inaccuracy of orbital information plays a large role for these sources, since their \gls{gw} signals are in the data stream for the entire mission duration. \Glspl{emri} are similarly long-lived sources that might be observable over the mission lifetime, or a significant fraction. The signals from such sources are expected to have a much higher complexity than that of Galactic-binary signals, or indeed any other source in the \gls{lisa} catalog. Nevertheless, a Galactic-binary signal essentially resembles that from a particular (albeit unrealistic) type of \gls{emri}: an early-stage quasi-circular \gls{emri}, where a single, slowly evolving, harmonic mode dominates the signal. Therefore, we conjecture that the findings presented here on orbital information accuracy will, at best, similarly apply to \glspl{emri}---rendering equal-armlength approximations to response functions unusable in parameter estimation for high-\gls{snr} sources. At worst, the bias incurred from even full treatments of the orbit and response function might be compounded by the higher waveform complexity, and might end up as the dominant source of error for these precisely modeled sources.

Massive black-hole binaries are going to be observed for much shorter stretches of time than Galactic binaries (a few days or weeks)~\cite[e.g.,][]{Katz:2020hku}. They will also be observed at higher \glspl{snr}. When analyzing massive black-hole binaries, due to their short-duration signal, the equal-armlength orbital parameters can be optimized (see \cref{sec:orbitaltraj}) to a smaller number of time points, potentially making this approximation more accurate. With that said, in a full parameter estimation or global fit setting, it would be unwise to assume different underlying orbits for different sources. Therefore, we categorize this technique as another potential option when \textit{searching} for massive black-hole binaries. Given fixed underlying orbits, we draw the conclusion that equal-armlength orbits for massive black-hole binaries are also not good enough for inference, viewing our Galactic binary example once again as conservative. The higher \glspl{snr} of massive black-hole binaries will risk larger bias, and may be more affected by instantaneous differences between accurate and inaccurate orbits if their mergers lie during a small time segment where the orbits are less similar.

In a more realistic analysis setup, the parameters of various overlapping sources must be estimated simultaneously by a \textit{global fit} technique. One can reasonably think that orbital parameters will be be included in this global fit, and therefore argue that the findings presented do not apply. However, if one uses a simple orbital model, such as the equal-armlength configuration, the fitted orbital parameters for the simple model will be, at best, identical to the parameters fitted to the true orbits, i.e., the exact same parameters used in this conservative study. Therefore, we believe that the results presented here remain valid in a global fit analysis.

Using a more accurate response function for parameter estimation, and in particular a more realistic orbital model, as we advocate here, requires some knowledge of the positions of the \gls{lisa} spacecraft, as well as the light travel times along the 6 constellation links. The light travel times can be computed from the spacecraft position, velocity and acceleration vectors~\cite[e.g., ][]{Hees:2014lsa}. The \gls{lisa} spacecraft follow the free-falling test masses they host, such that their trajectories are, to great precision, simple geodesics in the Solar System. These geodesics are easily computed using standard ephemerides and some initial conditions (this is how \gls{esa} orbits used in this paper have been calculated). The initial conditions are provided by on-ground estimation of absolute positions and velocities of the \gls{lisa} satellites by \gls{esa}'s ESTRACK system. Ground-tracking position estimates have $\sim \SI{10}{\kilo\meter}$ precision~\cite{mj2021}; this is several orders of magnitude below the difference between the fitted equal-armlength orbits and \gls{esa} orbits used in this study (c.f., \cref{fig:orbit_plot}). Therefore, we are confident that spacecraft trajectories will be known with sufficient accuracy not to be a limiting factor in source parameter recovery.

\section{Conclusion}
\label{sec:conclude}

In this article, we presented a new open-source tool for computing the general \gls{lisa} \gls{tdi} response to \glspl{gw} in the time domain, for arbitrary orbits and waveforms. The code is flexible and can be run on both \gls{cpu} and \gls{gpu}; however, a specialized implementation of the response function for \glspl{gpu} makes it fast enough ($\sim \SI{10}{\milli\second}$ for 4~years worth of data) to use in stochastic sampling algorithms, such as \glspl{mcmc}.

We took advantage of this newly-found speed to quantify the parameter-estimation bias, under current \gls{lisa} mission settings, incurred by analysis codes  that adopt equal-armlength \gls{lisa} orbits for the sake of computational efficiency. We found that the resulting waveform templates produce unacceptable bias in source frequency and position for loud Galactic-binary waveforms; however the corresponding matched-filtering \gls{snr} would still be appropriate to identify sources during search. 
We also tested hybrid templates that use accurate orbits for \gls{gw} projection, but equal armlengths for \gls{tdi}. Such templates could be implemented with straightforward modifications to equal-armlength codes maintaining their efficiency. These hybrid templates enable accurate parameter recovery except for the loudest sources (\gls{snr} $\sim 500$), where they produce significant bias.

These results indicate the importance of including accurate orbits in the \gls{lisa} signal model, and they highlight the need to upgrade current approximate methods in preparation for real data.
We expect that these conclusions would be strengthened for \gls{gw} signals more complex than the Galactic binaries studied here, such as \glspl{emri} or massive black-hole binaries. However, more studies must be carried out to obtain a definitive answer on the impact of orbital approximations for these types of sources. Other features of real data, such as nonstationary noise, gaps, glitches, etc., would also impact data-analysis performance, but we do not expect them to affect the general findings presented here.

\acknowledgments
J.-B.B. was supported by a NASA postdoctoral fellowship administered by USRA. A.J.K.C. acknowledges support from the NASA LISA Preparatory Science grant 20-LPS20-0005. M.V. was supported by the NASA LISA study office. This research was supported in part through the computational resources and staff contributions provided for the Quest/Grail high performance computing facility at Northwestern University. This paper also employed use of \texttt{SciPy}~\cite{2020SciPy-NMeth} and \texttt{Matplotlib}~\cite{Matplotlib}. Part of this research was carried out at the Jet Propulsion Laboratory, California Institute of Technology, under a contract with the National Aeronautics and Space Administration (80NM0018D0004). Copyright 2022. All rights reserved.

\bibliography{references}

\end{document}